# Explainable anomaly detection for sound spectrograms using pooling statistics with quantile differences

Nicolas Thewes[a], Philipp Steinhauer[b], Patrick Trampert[a], Markus Pauly[b,c] and Georg Schneider[a]

[a] Artificial Intelligence Lab, ZF Friedrichshafen AG, Scheer Tower II, Uni-Campus Nord, Geb. D5 2 66123 Saarbrücken, Germany

[b] Department of Statistics, TU Dortmund University, 44227 Dortmund

[c] Research Center for Trustworthy Data Science and Security, UA Ruhr, 44227 Dortmund

## 1 Introduction

Anomaly detection is the task of identifying rarely occurring (i.e. anormal or anomalous) samples that differ from almost all other samples in a dataset. As the patterns of anormal samples are usually not known a priori, this task is highly challenging. Consequently, anomaly detection lies between semi- and unsupervised learning.

The detection of anomalies in sound data, often called 'ASD' (Anomalous Sound Detection), is a sub-field that deals with the identification of new and yet unknown effects in acoustic recordings. It is of great importance for various applications in Industry 4.0. Here, vibrational or acoustic data are typically obtained from standard sensor signals used for predictive maintenance. Examples cover machine condition monitoring or quality assurance to track the state of components or products.

However, the use of intelligent algorithms remains a controversial topic. Management generally aims for cost-reduction and automation, while quality and maintenance experts emphasize the need for human expertise and comprehensible solutions. The importance of sound data for industrial applications is also reflected in the current literature. For example, the specifically designed MIMII Dataset (Sound Dataset for Malfunctioning Industrial Machine Investigation and Inspection) serves as a benchmark for sound-based anomaly detection [1]. Moreover, the DCASE (Detection and Classification of Acoustic Scenes and Events) community offers ASD challenges since 2016, see e.g. [7], [8],[21].

In this work, we present an anomaly detection approach specifically designed for spectrograms. The approach is based on statistical evaluations and is theoretically motivated. In addition, it features intrinsic explainability, making it particular suitable for applications in industrial settings.
Thus, this algorithm is of relevance for applications in which black-box algorithms are unwanted or unsuitable.

## 2 Related Work

The detection of anomalies is a broad, extensively researched field. A general introduction to outlier detection can be found in [2], [3] or [4]. In recent years, deep learning techniques are more and more applied in the field of anomaly detection by making use of the intrinsic feature-learning ability of deep learning methods, see e.g. [5] and [6] for a more specific review. In the following, we provide a short overview on publications in the ASD field. In contrast to our approach, most methods that have been applied for ASD are based on machine learning techniques, as the results of the DCASE challenges of the last two years [7], [8] show.

Autoencoder (AE) architectures are a frequently applied deep learning technique for ASD, see, e.g. [9], [10], [11], [12], [13] or [14]. They are used in different flavors. In two recent publications Coelho et. al. [9], [10] compare different AE architectures, like dense, CNN, and LSTM, on machine condition monitoring as well as on a cough classification task in an in-vehicle setting. Their results show that combining the three architectures leads to enhanced results when mel-spectrogram sound pre-processing was applied. Kapka [11] proposed an AE with an additional affine transformation of the latent vector based on some metadata classification of a sample, e.g. the ID of an industrial machine. Using reconstruction error as anomaly score on the DCASE 2020 Task 2 dataset, the author shows that this addition, on average, leads to an improved anomaly detection performance as compared to standard autoencoder approaches. Yet, the results vary strongly for different types of machines.

Although ASD is, by definition, an unsupervised learning problem, different approaches use so-called proxy outliers from related domains to improve differentiation between normal and anormal samples in the original domain, see [12], [13], [14], [15]. Koizumi et al. [12] use AE generated anomalous sounds to train another AE for anomaly detection and utilize the generated anomalies to increase the true-positive rate directly rather than implicitly when using only normal samples for autoencoder training. Kuroyanagi et al. in [13] use the DCASE 2020 Task 2 dataset to show that proxy-outliers that come from other machine IDs and machine types can be applied to perform anomaly detection on a specific machine ID. In their approach, which is like [16], they simultaneously train a neural network using cross-entropy loss for multiclass-classification as well as for producing compact representations in feature space for the normal class. This enables to produce features that improve anomaly detection by increasing inter-class distance for normal samples and anormal samples and simultaneously optimizing for low intra-class distance of normal samples in feature space. In a similar, but simpler, approach Primus et al. [14] use proxy outliers, which were carefully selected from DCASE 2020 Task 2 from different machines and machine IDs and only train a classification network on the normal samples and proxy-outliers. That way they achieve significant improvements over the baseline while showing that the selection of the 'correct' outliers is key.

Another approach, which is used by [15], is to use features from pretrained deep neural networks to perform anomaly detection using classic anomaly detection methods such as One-Class Support Vector Machines [18] or Isolation Forests [19]. Müller et al. [15] use the MIMII-Dataset [1] to show that features extracted from Neural Networks pretrained on ImageNet in combination with non-deep learning anomaly detection models lead to superior results as compared to an end-to-end trained AE. In particular, they find that ResNet [17] features in combination with Gaussian Mixture Models give the best results.

Finally, we want to mention an older approach, as it shares basic ideas with our work: Hazan et al. [20] use log-periodograms, obtain a mask by taking the maximum values of half of the training dataset and subsequently determine the values in the second half of the dataset that excess this mask. Those excess values are then modeled using a specific distribution. Finally, anomaly detection is performed by using the distribution of excess values to calculate the probability of normality of test samples.

# 3 Methods

## 3.1 Intuition

A spectrogram is basically an ordered matrix, where each entry gives the energy (or amplitude) of a certain frequency in a certain time interval. Hence, if a standardized measurement protocol is applied, different spectrograms can be compared entry-wise. A simple anomaly detection approach could be as follows:

1. Assume a single entry is anormal if its value is higher than expected, i.e., higher than most values at that frequency/time point.
2. For each entry calculate statistical measures based on a training data set, which contains ideally only normal samples. Examples are quantiles or mean values and standard deviations. The result of this step is a matrix with the same dimension as the training data spectrograms, which we refer to as reference-spectrogram, see Figure 1 a).
3. Subsequently, an entry-wise anomaly detection can be applied using similar heuristics as in Six Sigma one-sided process control charts [25], such as: 'values outside the *99%*-quantile are anormal', compare Figure 1 b).

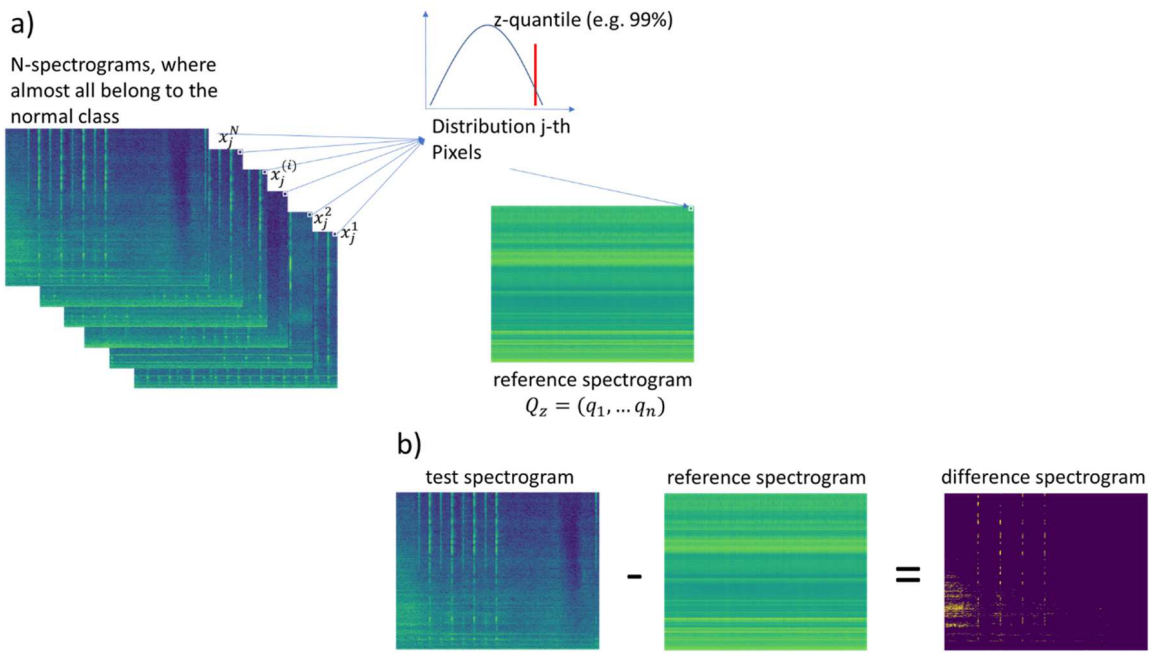

*Figure 1: Schematic explanation of our proposed approach. a) Given many spectrograms, where almost all belong to the normal class a reference-spectrogram can be calculated by using 'entry-wise' statistics, e.g. a 99% quantile. b) Based on the calculated reference, abnormality can be accessed on a single entry-level, by detecting the entries that have a higher value than their corresponding entry in the reference-spectrogram, e.g. by taking the difference between test spectrogram and reference.*

By adding the assumption of entry-wise independence a theoretically founded anomaly detection algorithm can be deduced.

## 3.2 Theoretical foundations

In this section a pre-stage of the anomaly detection algorithm is described to highlight the theoretical foundation of our approach.

Assume a set of N spectrograms that belong (mostly) to the normal class. Further, each spectrogram consists of *n* entries. Let $x_j^{(i)}$ denote the j-th entry of the i-th spectrogram with *i = 1, …, N* and *j = 1, …, n*. Then for every entry j the z-quantile is calculated over all spectrograms N, e.g. from a training set. This results in

$$q_j := z - quantile\ of\ \left(x_j^{(1)}, \ldots, x_j^{(N)}\right)\ with\ z \in [0,1],\ j = 1, \ldots, n.$$

These *n* quantiles make up the reference spectrogram denoted as $Q_z = (q_1, \ldots, q_n)$.

Consider a randomly selected spectrogram of a test set. Using the entry-wise anomaly detection based on a z-quantile reference spectrogram outlined in the previous section, the probability of any entry in the test spectrogram to have a higher value than the respective quantile is 1-z. Assuming entry-wise independence and given the spectrograms under inspection have *n* entries in total, the number *k* of entries lying above their respective quantile follows a binomial distribution. With that, the probability that a test spectrogram does *not* belong to the training set distribution, which means it is an anomaly, can be calculated by counting the number of entries that lie outside their corresponding *z*-quantile:

$$P_{anormal}(k) = \binom{n}{k} \cdot (1-z)^k \cdot z^{(n-k)}$$

This approach already works reasonably well but one simple extension can make it much more effective. The binomial distribution approach does not consider by how much a certain entry rises above its *z*-quantile. However, the absolute height of a value can be of great importance when trying to find anormal spectrograms. A small number of large deviations might be much more suspicious than many small deviations, especially in the presence of random noise.

### 3.3 Proposed algorithm

Let $(w_1, \ldots w_n)$ be the entries of a test spectrogram W. Given a reference spectrogram Q that was calculated as described above, we define the difference spectrogram $D_W$ as:

$$\begin{aligned} D_W &= max(0, W - Q) \\ &= \left(max(0, w_1 - q_1), max(0, w_2 - q_2), \ldots, max(0, w_n - q_n)\right) \\ &= (d_1, d_2, \ldots, d_n). \end{aligned}$$

The difference spectrogram can then be used to calculate an anomaly score $a_W$ of W with different metrics. In this work we propose four deviation metrics:

- Counting – The anomaly score is represented by counting how many entries k in the difference spectrogram D are larger than zero: $a = k := \sum_{i=1}^{n} I_{\{d_i > 0\}}$.
- Sum – Sum over all entries larger than zero: $a = \sum_{i=1}^{n} d_i \cdot I_{\{d_i > 0\}}$.
- Mean – Mean of all entries that are larger than zero: $a = \frac{1}{k}\sum_{i=1}^{n} d_i \cdot I_{\{d_i > 0\}}$.
- Binomial – Using binomial distribution: $a = \binom{n}{k} \cdot (1-z)^k \cdot z^{(n-k)}$.

To sum up, we introduce a new anomaly detection model for sound spectrograms that is based on an entry-wise pooling via quantiles with two hyperparameters: The quantile used for pooling and the metric used to calculate the deviation between a reference and a test spectrogram. In the next section we evaluate the performance of the proposed algorithm using the MIMII dataset [1].

# 4 Experiments

## 4.1 Dataset

To test the performance of our proposed algorithm, we use the MIMII dataset [1]. This dataset consists of normal and anormal sounds of four different machine types (fan, pump, valve and slider/slide rail) with four individual machine IDs (0,2,4,6). Data was recorded using a microphone array of eight microphones resulting in eight channel recordings where each recording has a length of 10 seconds sampled with 16 kHz. To better reflect the intrinsic challenges of an industrial setting additional background noise was captured and mixed with the sound recordings using different signal-to-noise ratios (-6 dB, 0 dB, 6 dB). We start with the 0 dB data for comparison to the MIMII baseline, then we show the results of experiments with the three different noise levels and end by comparing our results on the -6 dB data with the anomaly detection procedure recently published in [15]. In the following we give a detailed description of our preprocessing, hyperparameter tuning, and evaluation.

## 4.2 Preprocessing

Each sample of the MIMII dataset is provided as a WAVE file. To transform the acoustic data into frequency space we use the python package *librosa* [22]. Spectrograms are calculated using *short-term-Fourier-transformation* by using the librosa '*stft*' method with a window size of 2048, window type 'hann', and constant padding. Subsequently, spectrograms are converted to DB-scale using the

'amplitude_to_db' method. For the following steps we save the spectrograms as *numpy arrays* [23]. Sample normal and anormal spectrograms for two machines are displayed in Figure 2.

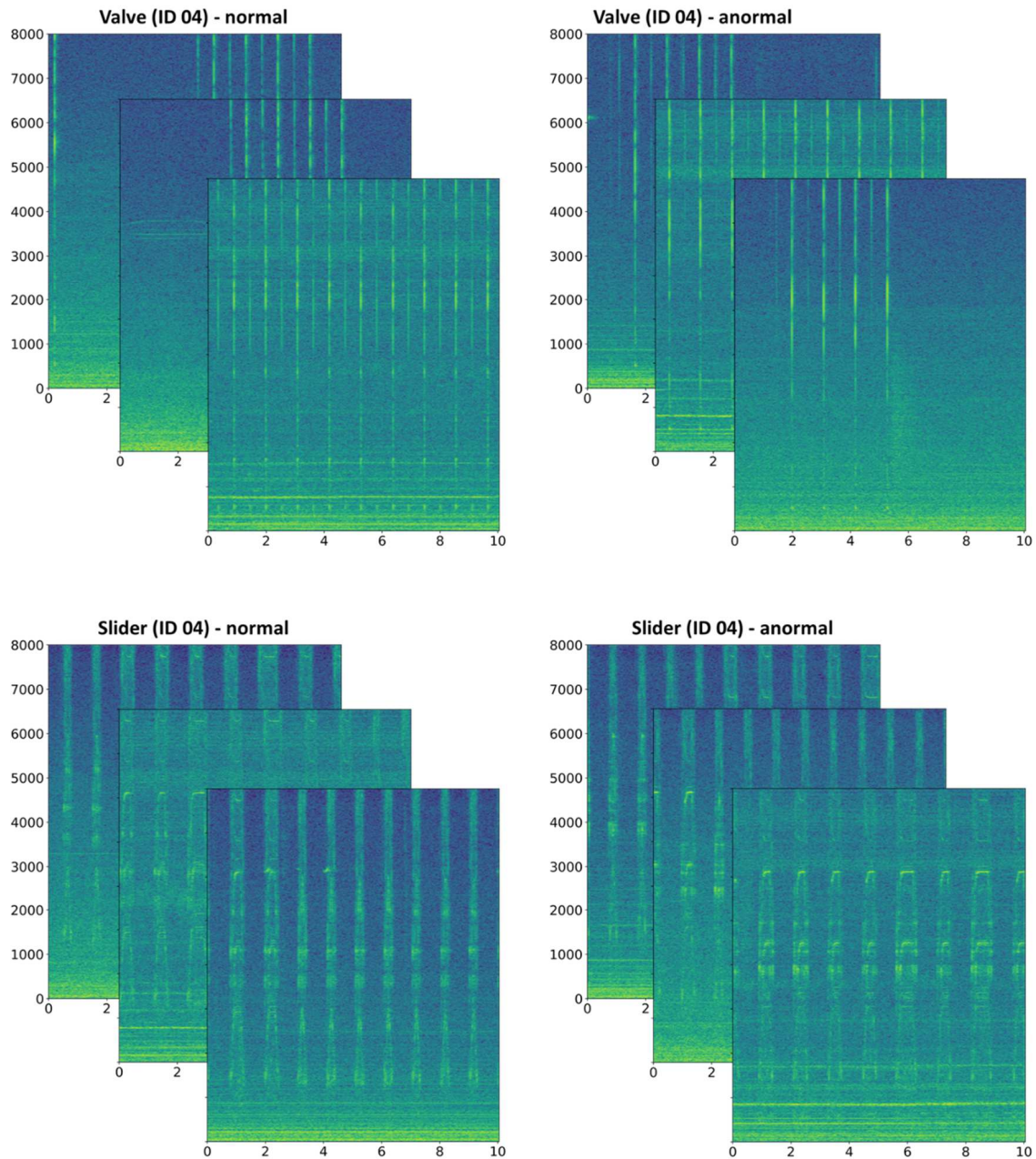

*Figure 2: Sample spectrograms calculated as described in 4.2. For two machines (type/ID) three normal and anormal spectrograms are displayed, the y-axis displays the frequency axis ranging from 0 to 8 kHz, the x-axis displays the measurement time ranging from 0 to 10 s.*

## 4.3 Evaluation

To get results that are comparable with other results in the literature, we use *area under the curve* (AUC) of the *receiver operating characteristic* (ROC), abbreviated AUC in the following. This metric is standard for the evaluation of anomaly detection approaches. AUC was also used to score the baseline algorithm that has been published alongside the dataset by the authors of the MIMII dataset. In short, AUC evaluates how well an anomaly score can separate normal samples from anormal samples [24].

To tune hyperparameters and get an unbiased estimate of the final performance we split the data of each machine ID into training, validation for hyperparameter estimation, and test set by

1. splitting the anormal samples by 50 percent for validation 50 percent for test and
2. adding the same number of normal samples to each of the anormal sample portions.

That way we get two separate balanced datasets for hyperparameter tuning and final testing. This procedure is repeated five times using different random seeds. For each combination of training, validation, and test data, we tune hyperparameters on the validation set and subsequently calculate the test set AUC. The final estimate is then calculated by averaging the five test set estimates per machine ID. Table 1 summarizes the number of normal and anormal samples for two machines.

| Machine type | Machine ID | Split | #normal | #anormal |
| --- | --- | --- | --- | --- |
| Fan | 0 | Training | 604 | 0 |
| Fan | 0 | Validation | 203 | 203 |
| Fan | 0 | Test | 204 | 204 |
| Slide rail | 4 | Training | 356 | 0 |
| Slide rail | 4 | Validation | 89 | 89 |
| Slide rail | 4 | Test | 89 | 89 |

*Table 1: Number of normal and anormal samples per data split for two exemplary machine types/IDs. For a full view on the number of samples, see [1].*

### 4.4 Hyperparameter-Tuning

Our algorithmic approach features two hyperparameters. The first one is the method to calculate the reference spectrogram from the training set spectrograms. Here we use quantiles as distribution indicators, thus, we restrict our search to the quantile z-value and search through the following values: 0.5, 0.6, 0.7, 0.8, 0.9, 0.95, 0.99. The second hyperparameter is the metric used to calculate the deviation score based on a difference spectrogram. Possible choices in this case are Counting, Sum, Mean, and Binomial. We perform a grid search to find the optimal hyperparameter combination. Table 2 shows representative results of the hyperparameter tuning, i.e. the best AUC on the respective validation dataset, for two machines and the corresponding five splits.

| Machine type | Machine ID | # split | z-quantile | deviation-metric | AUC on validation |
|---|---|---|---|---|---|
| Slide rail | 4 | 1 | 0.99 | Mean | 1.0 |
| Slide rail | 4 | 2 | 0.95 | Mean | 1.0 |
| Slide rail | 4 | 3 | 0.95 | Mean | 1.0 |
| Slide rail | 4 | 4 | 0.95 | Mean | 1.0 |
| Slide rail | 4 | 5 | 0.99 | Mean | 1.0 |
| Fan | 6 | 1 | 0.9 | Mean | 0.65 |
| Fan | 6 | 2 | 0.9 | Mean | 0.59 |
| Fan | 6 | 3 | 0.995 | Mean | 0.61 |
| Fan | 6 | 4 | 0.9 | Mean | 0.61 |
| Fan | 6 | 5 | 0.995 | Sum | 0.62 |

*Table 2: Exemplary results of hyperparameter tuning for two machines. For each split we show the hyperparameter combination that results in the best validation AUC.*

Based on our hyperparameter tuning, we can state that for the MIMII dataset our model favors quantiles above 0.9 and "mean" as deviation metric.

In the next section we compare our results against the baseline that has been published alongside the MIMII dataset.

# 5 Results

In the following subsections we present and discuss our results and compare them to the AE based baseline published alongside the MIMII dataset [1]. At that point we want to emphasize that our approach is setup directly on sound spectrograms. These are the output of a fast-Fourier transformation without further processing as described in Section 4.2.

## 5.1 Test set results

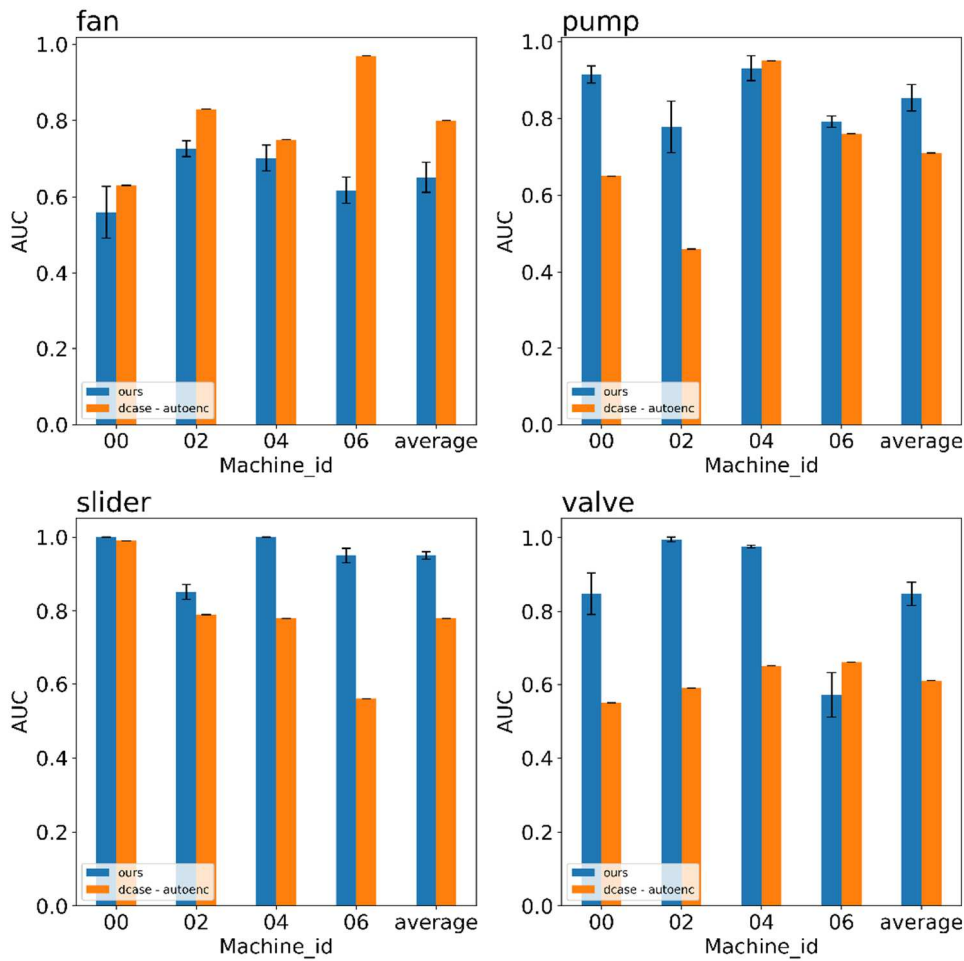

*Figure 3: AUC results for our approach and the autoencoder baseline approach for the four different machine types (fan, pump, valve and slider).*

Figure 3 shows that based on our experiments our approach is on par or better than the provided baseline in 12 out of 16 machines based on AUC values. Averaged on machine level, we surpass the baseline approach in 3 out of 4 cases. For machine ID 2 and 4 of valve and machine ID 0 and 4 of slide rail our model almost perfectly separates normal and anormal cases. Interestingly, for machine type fan the performance of our model is worse than the baseline in most cases and worse compared to all other machine types.

In the next section we show that our model can intrinsically produce an explainability output, which can be used to understand anomaly properties on a single sample level. With this we take a closer look on the results of our model for machine ID 2 of machine type valve and machine ID 6 of machine type fan to provide further insights towards benefits and limitations of our approach.

## 5.2 Explainability

As described in Section 3.3 our approach is based on calculating the difference between a reference spectrogram Q and a test spectrogram *W*. Thereby, the resulting difference *D* is a direct visualization of the positions in the test spectrogram, usually energies or amplitudes, that surpass the value of – in the case of a 99%-quantile as pooling statistics – 99% of the values of the training spectrograms at that exact position. Figure 4 shows an anormal and a normal spectrogram alongside its explainability output. The explainability output of the normal spectrogram in the lower row shows rather randomly distributed points. Yet, the explainability output of the anormal spectrogram in the upper row shows higher excitations on different points that resemble structures in the original spectrogram. Firstly, there are significant bright spots at 4 kHz that show up periodically, secondly, we find straight horizontal excitations below 1 kHz. For an engineer familiar with the machine at hand, those findings can be extremely valuable to identify the root cause of the current problem. Additionally, the explainability output can be used as input for downstream classification models.

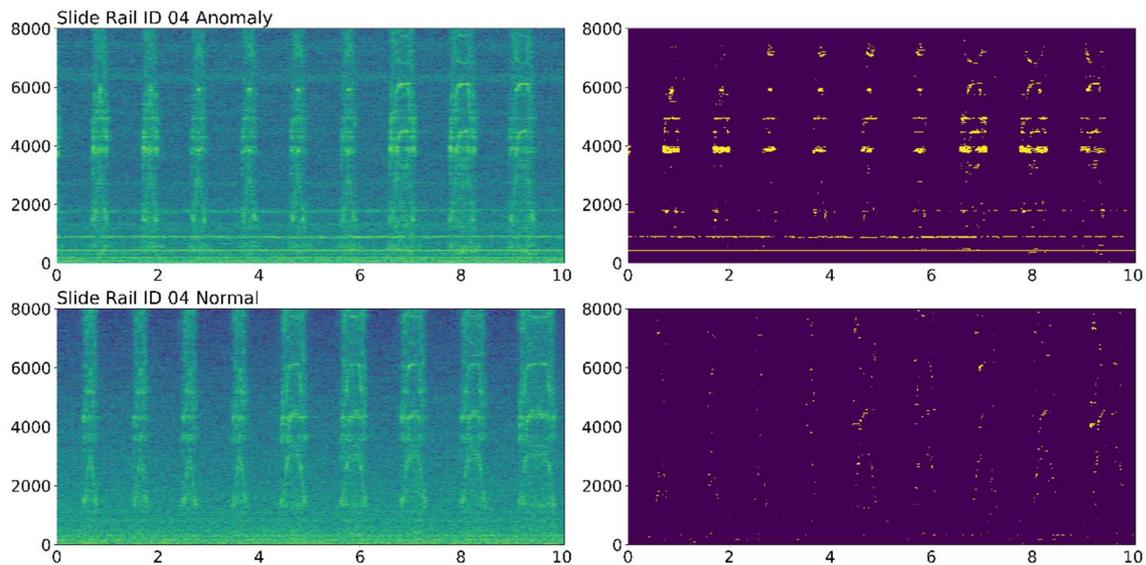

*Figure 4: Exemplary explainability spectrograms (right) and the corresponding original spectrograms (left). Upper row shows an anormal spectrogram, lower row a normal one. This example is from slider ID 4 and the explainability output was generated using a 99%-quantile approach.*

## 5.3 In-depth evaluation of results

To gain a deeper understanding of our approach and its applicability we perform an in-depth analysis of the valve ID 2 and fan ID 6 results. We chose those as they represent cases where our model outperforms the baseline (valve ID 2) and vice versa (fan ID 6).

### 5.3.1 Valve ID 2

For machine type valve and machine ID 2, our approach generates an anomaly score that induces an almost perfect split between normal and anormal samples, the AUC is 0.994(7). Figure 5 shows three randomly chosen anormal spectrograms on the left and three randomly chosen normal spectrograms from a test set alongside their explainability output. While it is hard to see a significant difference between the two classes by eye in the spectrograms, the explainability output clearly shows that the vertical lines in the spectrograms, which may relate to opening and closing the valve, show significantly higher excitations for the anormal spectrograms.

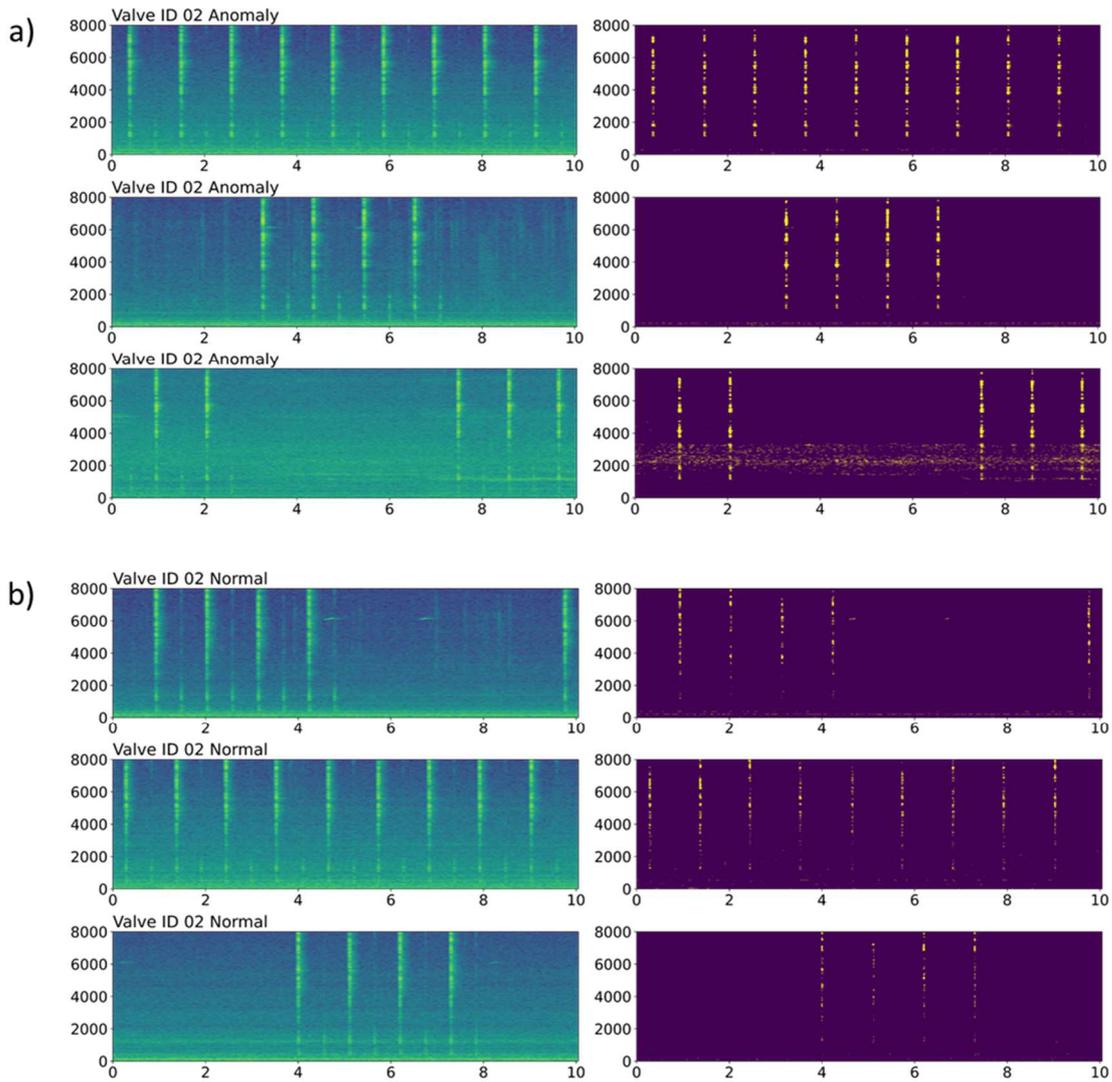

*Figure 5: Three anormal spectrograms (a) and three normal spectrograms alongside (b) their explainability output, each chosen randomly from a test set. The spectrograms belong the machine type valve and machine ID 2.*

Next, we display three normal spectrograms that have been assigned a relatively high anomaly score, cf. Figure 6. The vertical lines visible in the explainability output doesn't seem to be higher than in the case of the randomly chosen normal spectrograms in Figure 5, however, in addition horizontal lines are visible over the whole measurement time. At that point, a human expert could enter the loop to identify if this excitation is the result of an anormal behaving machine or other non-failure indicating reasons like background noise.

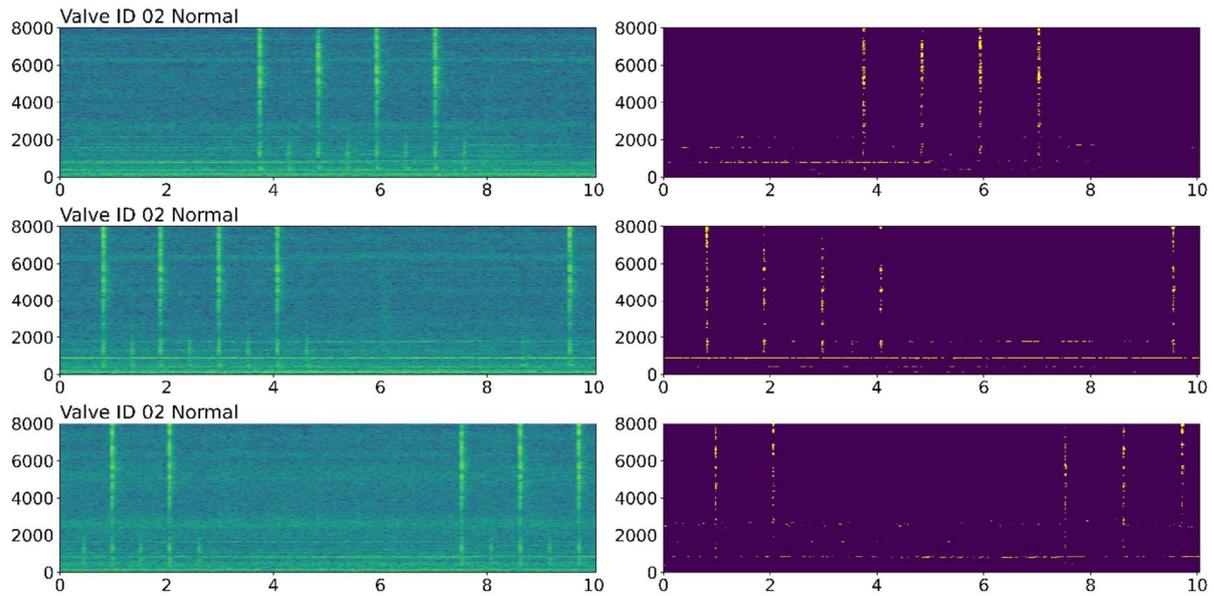

*Figure 6: Three spectrograms of class normal (left) that have been assigned high anomaly scores with their respective explainability spectrograms (right). The spectrograms belong to machine type valve with machine ID 2.*

### 5.3.2 Fan ID 6

For machine type fan and machine ID 6 our approach achieves an AUC of 0.62(3) as compared to 0.97 with the baseline model. Hence, the AE approach can learn patterns that are largely inaccessible to our statistics-based approach. Figure 7 shows three normal and anormal samples with high anomaly scores. For each class the scores are in the same range. For the normal samples, it is striking that the excitations highlighted by our approach that led to the high anomaly score, show almost no specific structure but are rather distributed across the whole spectrogram. The same is true for the last one of the anormal samples and most other anormal spectrogram, which are not shown here. Figure 8 shows the three spectrograms with the lowest anomaly score for normal and anormal class. Again, for both classes the score is in the same range and no obvious difference between normal and anormal samples can be identified by eye, neither in the original spectrograms nor in the explanation outputs.

A detailed investigation of the differences between normal and anormal samples for that machine, which would also include human experts, could help to understand the weak performance of our model but is beyond the scope of this study.

Basically, as our approach intrinsically depends on the assumption that anormal spectrograms contain excitations *higher* than what is normally observed. A possible explanation could be that, for this machine, anomalies are caused by excitations that are too low rather than too high or due to other features, which we cannot (yet) include in our model. Expanding our approach in these directions is a future task.

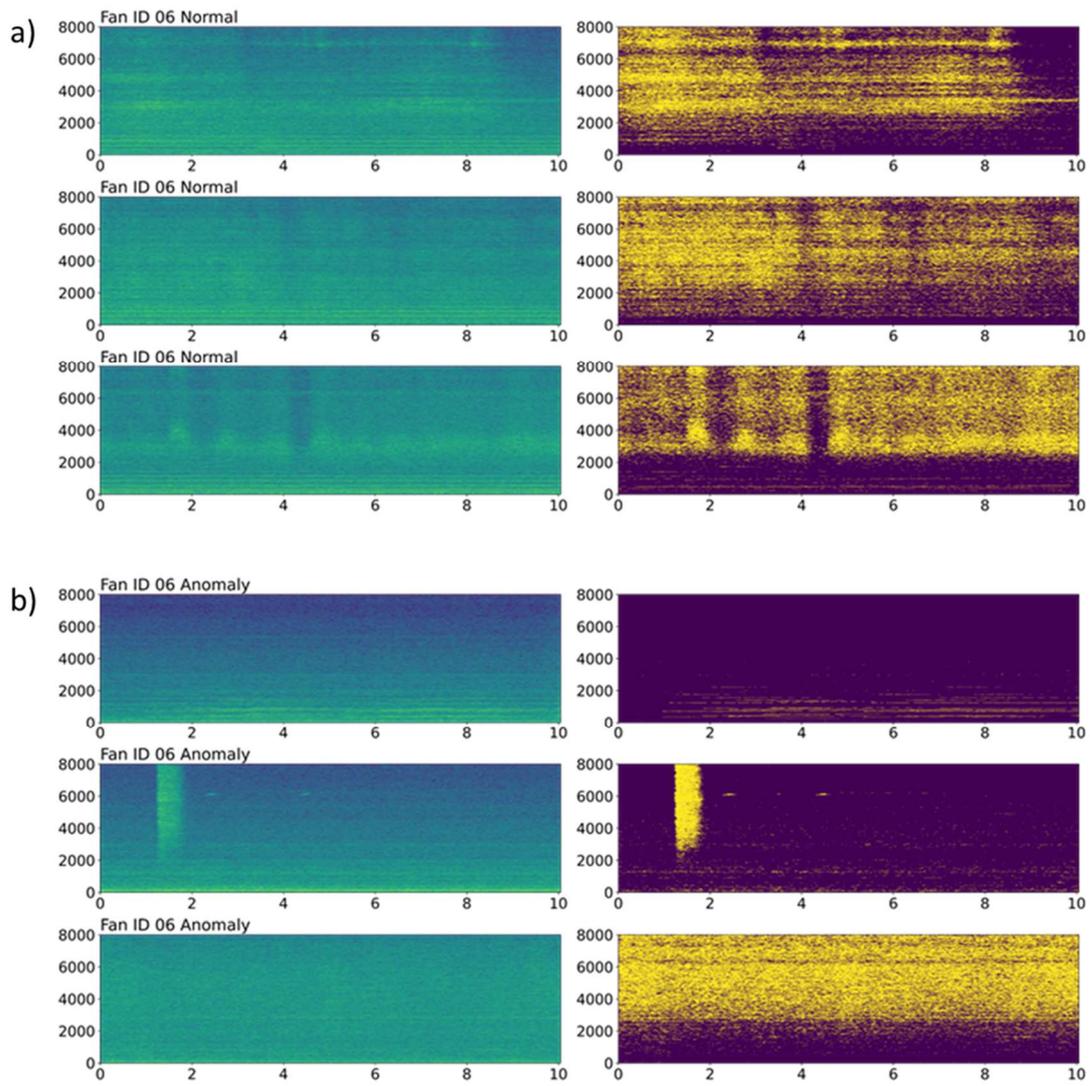

*Figure 7: Normal spectrograms with high anomaly scores (a) and anormal ones with high anomaly scores (b). Samples belong to machine type fan and machine ID 6.*

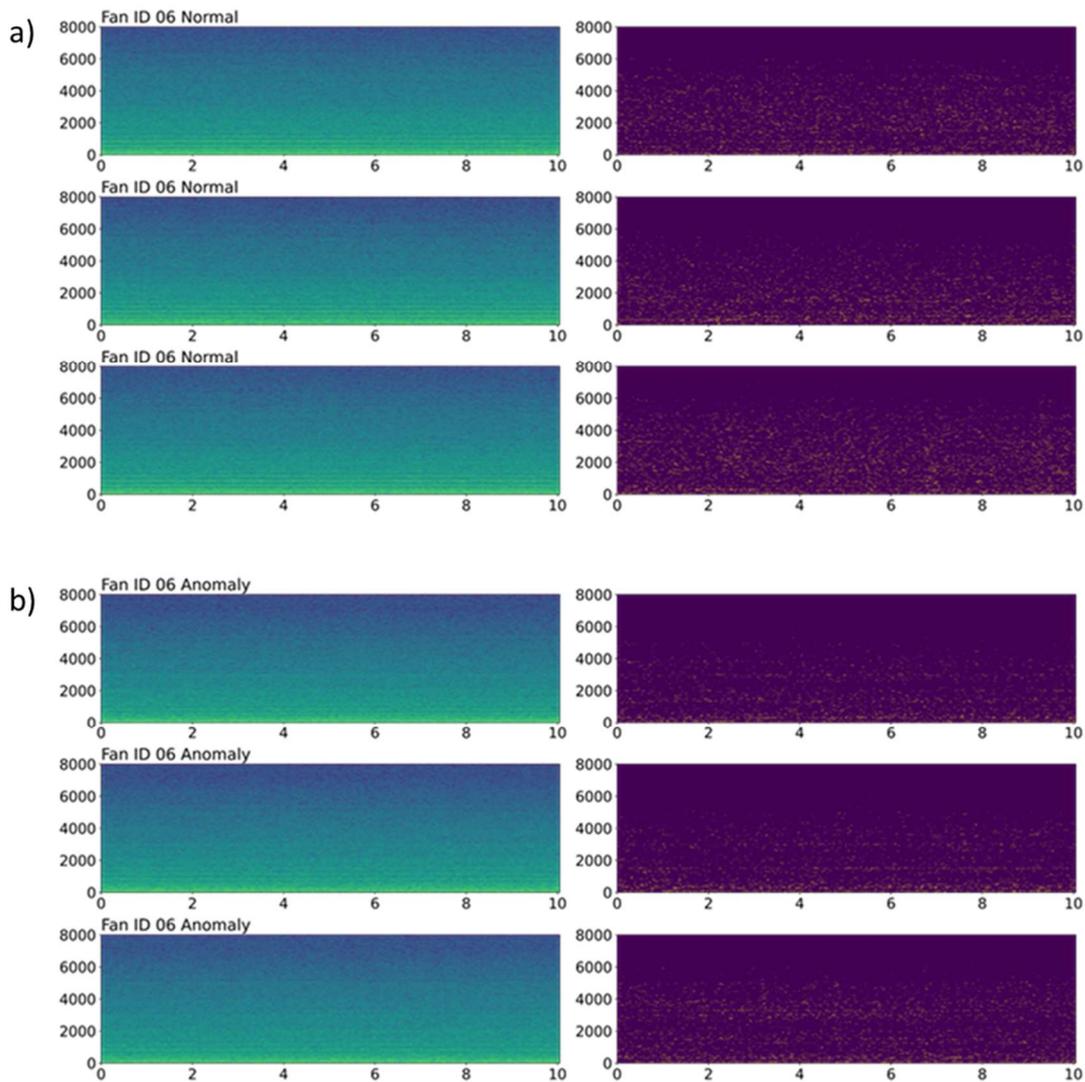

*Figure 8: As Figure 7 but with normal (a) and anormal (b) spectrograms that our approach assigns low anomaly scores, too.*

### 5.4 Influence of noise

In industrial settings, noise is an omnipresent source of difficulties when it comes to dealing with sound data. For the task of anomaly detection, noisy environments can lead to false positives or might mask patterns that are important for anomaly detection models. Hence, results of algorithmic anomaly detection are usually worse the noisier sound recordings are [1]. The MIMII-dataset provides sound recordings with three different noise levels, -6 dB, 0 dB and 6 dB. Figure 9 shows the results of the anomaly detection with our algorithm on all three noise levels of the MIMII dataset. In general, we see decreasing detection performance for increasing noise levels. However, the severity of this effect varies strongly for different machines, as for example we see a huge influence of the noise level for machine type fan, whereas almost no influence can be measured for machine type slider.

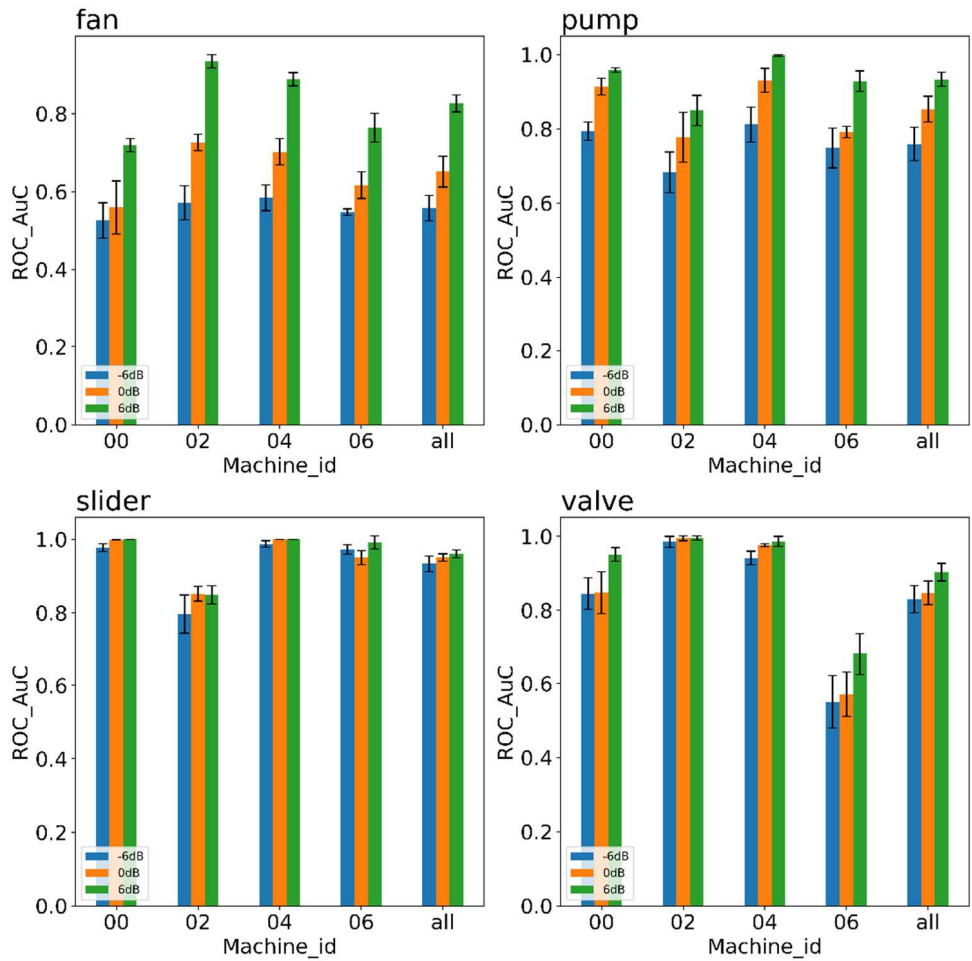

*Figure 9: Results of our anomaly detection algorithm for different noise levels of the MIMII dataset.*

To provide further placement in recent literature, we compare our results on the -6 dB data against the approach recently published by R. Müller et. al. [15]. They use state-of-art CNN architectures pretrained on image net to calculate feature vectors that subsequently serve as input to classic anomaly detection algorithms like One-Class Support vector machine, Kernel Density Estimation, and the like. In Figure 10 we compare the results of our algorithm to the best and second-best result of the paper "Acoustic Anomaly Detection for Machine Sounds based on Image Transfer Learning" [15]. In general, the results of our model and of the Image-Transfer-Learning (ITL) approach [15] vary strongly for different machines. As for the 0 dB data, our approach performs weaker than the deep learning-based approach for machine type fan. For the other three machine types, our approach is on a par for pump and performs better on slider and valve than the ITL approach.

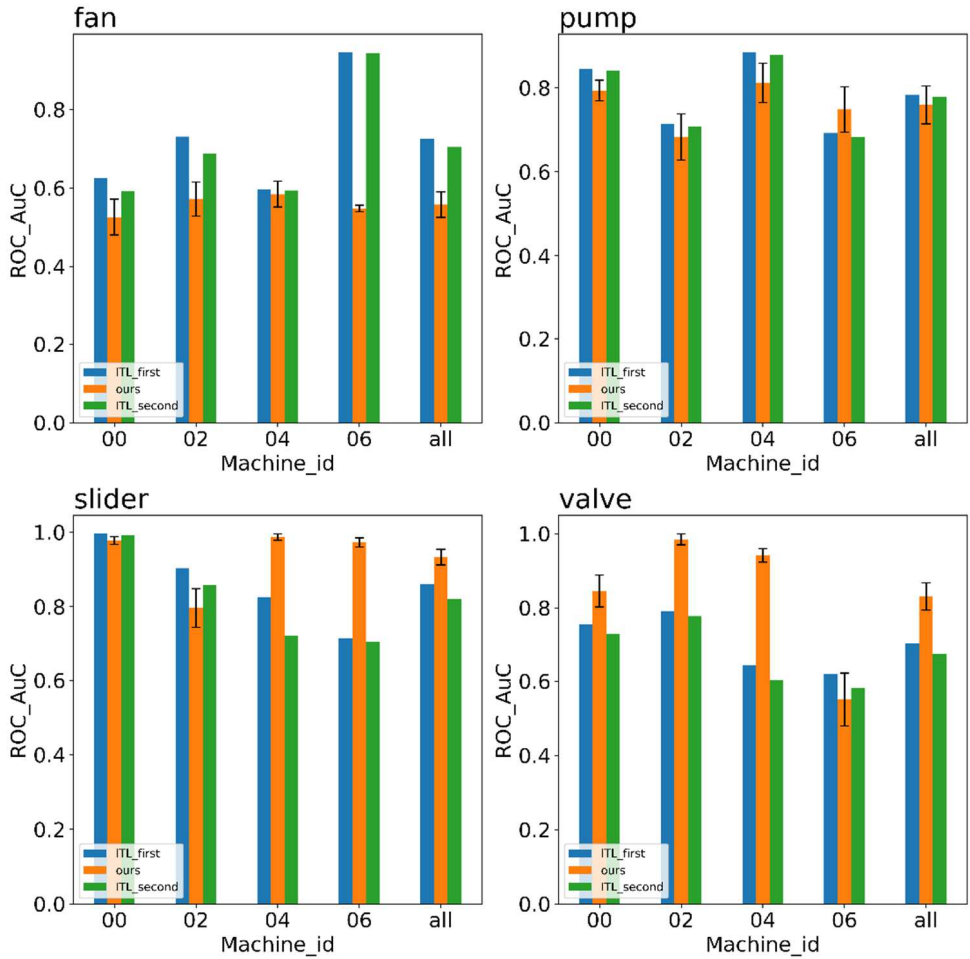

*Figure 10: Comparison of our approach and the approach published in the paper „Acoustic Anomaly Detection for Machine Sounds based on Image Transfer Learning" (ITL)) [15]. Anomaly detection results are evaluated using AUC and experiments are performed using the -6 dB noise level of the MIMII dataset [1].*

# 6 Relation to binomial distribution

In Section 3.1. we outlined the intuition of our approach with relation to the binomial distribution. In this section we investigate this relation by comparing experimentally obtained deviation counts with the corresponding theory based on the binomial distribution. To obtain a statistically relevant number of samples we conduct this experiment using acoustic data provided by a ZF Friedrichshafen AG production side. More specifically we use data from acoustic testing of passenger car transmissions. Each spectrogram consists of N=56942 'pixels'.

We perform the following steps for five different train/test splits and average the results:

- Take 50000 spectrograms from acoustic testing.
- Split the spectrograms into training and test sets using an 80:20 ratio.
- Calculate a reference spectrogram as described in Section 3.3. using quantiles with z-values of 0.5, 0.75, 0.9, 0.95, and 0.99.
- For each spectrogram in the test set, count the number of 'pixels' exceeding the corresponding value from the reference.
- Calculate the mean of the deviation and compare to the binomial distribution, given by $(1-z) * N$.

In Figure 11, we present the relative deviation between the expected value and the experimentally observed means.

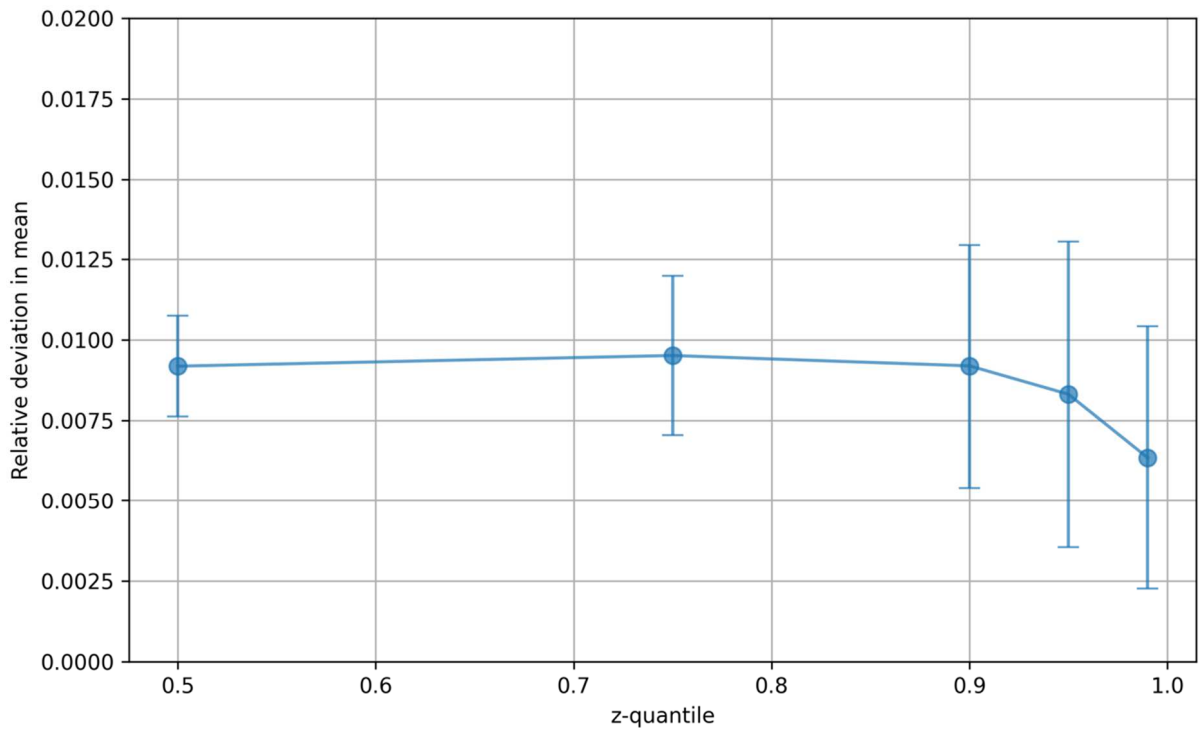

Figure 11: Relative deviation between the number of `pixel' larger than the corresponding reference value and the expected value based on binomial distribution. Each point denotes the mean and standard deviation of the relative deviation averaged over five different splits.

The experimentally determined values deviate from the expected values (based on binomial distribution) by slightly less than one percent for each z-quantile value considered. Interestingly we observe substantially larger standard deviations between experiments and theory (not shown here), with larger standard deviations for the experimental setting. We believe that this arises from the violation of the i.i.d. assumption in the experimental setting: neighboring frequencies and time windows are, of course, highly correlated. Incorpoarting theoretical approaches that account for the non-i.i.d. nature of observations in machine sound spectrograms will be an important future direction of our research.

# 7 Conclusion

In this work we presented a novel approach for identifying anormal spectrograms in machine sound data. Our approach is based on the entry-wise statistical evaluation of spectrograms and can be theoretically motivated. Additionally, the proposed algorithm offers intrinsic explainability making it particular suitable for applications that involve or require human supervision.

Applied to the MIMII dataset we achieved AUC scores of more than 0.8 for 11 out of 16 machines and more than 0.97 for four machines, which means an almost perfect split between normal and anormal samples. On average we outperformed the AE baseline for three out of four machine types. Interestingly, for machine type fan our model performed much weaker as compared to slide rail, pump, and valve.

Consistent with the baseline results, our approach is affected by noise levels with decreasing detection performance for increasing noise. However, this general pattern does not apply to every machine. For high noise levels of -6 dB, we compared the results of our model to the results of the deep learning-based approach published in [15]. Here, even though our approach is much less complex it achieves a detection performance comparable to that of [15].

At that point, we like to highlight two central observations anomalous sound detection:

1. The noisier the sound recordings are, the more difficult it becomes to detect anomalies. Hence, from a data-centric perspective, efforts should primarily focus on obtaining low noise recordings.
2. Anomaly detection performance varies vastly between different machine types and even individual machines of the same type (e.g., even machine IDs). Currently, there is no one-size-fits all method for ASD. Hence, in practical applications, the granularity, at which algorithms and hyperparameters are tuned and then applied must be determined experimentally, taking into account the specific technical conditions.

In summary, we propose a new algorithmic approach for anomalous sound detection that is theoretically founded, easy to implement, exhibits good to very good performance, and an intrinsic explainability. For three out of four machines in the MIMII dataset, the algorithm showed comparable or better detection performance even if compared to deep learning-based approaches. That is why the algorithm described in this work is used productively for detecting anormal behavior of ZF products during vibration quality tests, also called NVH tests. Those tests include human supervision and inspection. Therefore, and because the tested products are usually expensive, an algorithm that can explain why it assigned a high anomaly score to a product is of utmost importance. Additionally, we experience a higher acceptance of algorithms in production settings if the basic idea behind the algorithm is easily explainable also to non-machine learning experts. Furthermore, this can pave the way towards the usage of more complex and sometimes better performing machine learning algorithms at a later stage.

We believe that our algorithm has the potential to serve as standard starting point for anomalous sound detection in industrial applications. In the future, we plan to evaluate a two-sided version of our method and want to adopt more enhanced quantile regression as well as multivariate quantile-based approaches that account for the strong correlations between neighboring values in sound spectrograms.

# References

[1] H. Purohit et al., "MIMII Dataset: Sound Dataset for Malfunctioning Industrial Machine Investigation and Inspection," arXiv:1909.09347 [cs, eess, stat], Sep. 2019, Accessed: Dec. 21, 2021. [Online]. Available: http://arxiv.org/abs/1909.09347

[2] C. C. Aggarwal, "An Introduction to Outlier Analysis," in Outlier Analysis, C. C. Aggarwal, Ed. Cham: Springer International Publishing, 2017, pp. 1–34. doi: 10.1007/978-3-319-47578-3_1.

[3] V. Chandola, A. Banerjee, and V. Kumar, "Anomaly detection: A survey," ACM Comput. Surv., vol. 41, no. 3, p. 15:1-15:58, Jul. 2009, doi: 10.1145/1541880.1541882.

[4] M. A. F. Pimentel, D. A. Clifton, L. Clifton, and L. Tarassenko, "A review of novelty detection," Signal Processing, vol. 99, pp. 215–249, Jun. 2014, doi: 10.1016/j.sigpro.2013.12.026.

[5] R. Chalapathy and S. Chawla, "Deep Learning for Anomaly Detection: A Survey," arXiv:1901.03407 [cs, stat], Jan. 2019, Accessed: Jan. 15, 2021. [Online]. Available: http://arxiv.org/abs/1901.03407

[6] G. Pang, C. Shen, L. Cao, and A. V. D. Hengel, "Deep Learning for Anomaly Detection: A Review," ACM Comput. Surv., vol. 54, no. 2, p. 38:1-38:38, Mar. 2021, doi: 10.1145/3439950.

[7] N. Ono et al., Proceedings of the 5th Workshop on Detection and Classication of Acoustic Scenes and Events (DCASE 2020). Zenodo, 2020. doi: 10.5281/zenodo.4061782.

[8] F. Font, A. Mesaros, D. P. W. Ellis, E. Fonseca, M. Fuentes, and B. Elizalde, Proceedings of the 6th Workshop on Detection and Classication of Acoustic Scenes and Events (DCASE 2021). Barcelona, Spain: Music Technology Group - Universitat Pompeu Fabra, 2021. doi: 10.5281/zenodo.5770113.

[9] G. Coelho, L. M. Matos, P. J. Pereira, A. Ferreira, A. Pilastri, and P. Cortez, "Deep autoencoders for acoustic anomaly detection: experiments with working machine and in-vehicle audio," *Neural Comput & Applic*, May 2022, doi: 10.1007/s00521-022-07375-2.

[10] G. Coelho et al., "Deep Dense and Convolutional Autoencoders for Machine Acoustic Anomaly Detection," in Artificial Intelligence Applications and Innovations, Cham, 2021, pp. 337–348. doi: 10.1007/978-3-030-79150-6_27.

[11] S. Kapka, "ID-Conditioned Auto-Encoder for Unsupervised Anomaly Detection," arXiv:2007.05314 [cs, eess], Nov. 2020, doi: 10.5281/zenodo.4061782.

[12] Y. Koizumi, S. Saito, H. U. Y. Kawachi, and N. Harada, "Unsupervised Detection of Anomalous Sound based on Deep Learning and the Neyman-Pearson Lemma," IEEE/ACM Trans. Audio Speech Lang. Process., vol. 27, no. 1, pp. 212–224, Jan. 2019, doi: 10.1109/TASLP.2018.2877258.

[13] I. Kuroyanagi, T. Hayashi, K. Takeda, and T. Toda, "Anomalous Sound Detection Using a Binary Classification Model and Class Centroids," in 2021 29th European Signal Processing Conference (EUSIPCO), Aug. 2021, pp. 1995–1999. doi: 10.23919/EUSIPCO54536.2021.9616198.

[14] P. Primus, V. Haunschmid, P. Praher, and G. Widmer, "Anomalous Sound Detection as a Simple Binary Classification Problem with Careful Selection of Proxy Outlier Examples," arXiv:2011.02949 [cs, eess], Nov. 2020, Accessed: Oct. 18, 2021. [Online]. Available: http://arxiv.org/abs/2011.02949

[15] R. Müller, F. Ritz, S. Illium, and C. Linnhoff-Popien, "Acoustic Anomaly Detection for Machine Sounds based on Image Transfer Learning," Aug. 2022, pp. 49–56. Accessed: Aug. 22, 2022. [Online]. Available: https://www.scitepress.org/Link.aspx?doi=10.5220/0010185800490056

[16] P. Perera and V. M. Patel, "Learning Deep Features for One-Class Classification," IEEE Transactions on Image Processing, vol. 28, no. 11, pp. 5450–5463, Nov. 2019, doi: 10.1109/TIP.2019.2917862.

[17] K. He, X. Zhang, S. Ren, and J. Sun, "Deep Residual Learning for Image Recognition," 2016, pp. 770–778. Accessed: Aug. 23, 2022. [Online]. Available: https://openaccess.thecvf.com/content_cvpr_2016/html/He_Deep_Residual_Learning_CVPR_2016_paper.html

[18] B. Schölkopf, R. C. Williamson, A. Smola, J. Shawe-Taylor, and J. Platt, "Support Vector Method for Novelty Detection," in Advances in Neural Information Processing Systems, 1999, vol. 12. Accessed: Aug. 23, 2022. [Online]. Available: https://proceedings.neurips.cc/paper/1999/hash/8725fb777f25776ffa9076e44fcfd776-Abstract.html

[19] F. T. Liu, K. M. Ting, and Z.-H. Zhou, "Isolation Forest," in 2008 Eighth IEEE International Conference on Data Mining, Dec. 2008, pp. 413–422. doi: 10.1109/ICDM.2008.17.

[20] A. Hazan, J. Lacaille, and K. K. Madani, "Extreme value statistics for vibration spectra outlier detection," in International Conference on Condition Monitoring and Machinery Failure Prevention Technologies, Londres, United Kingdom, Jun. 2012, p. p.1. Accessed: Dec. 22, 2021. [Online]. Available: https://hal.archives-ouvertes.fr/hal-00681045

[21] "DCASE2020 Challenge - DCASE." https://dcase.community/challenge2020/ (accessed Sep. 19, 2022).

[22] B. McFee et al., "librosa: Audio and Music Signal Analysis in Python," Proceedings of the 14th Python in Science Conference, pp. 18–24, 2015, doi: 10.25080/Majora-7b98e3ed-003.

[23] C. R. Harris et al., "Array programming with NumPy," Nature, vol. 585, no. 7825, Art. no. 7825, Sep. 2020, doi: 10.1038/s41586-020-2649-2.

[24] Provost, Foster J., et al. The case against accuracy estimation for comparing induction algorithms. in: ICML. 1998. S. 445-453.

[25] (2000). SHEWHART CONTROL CHARTS . In: Swamidass, P.M. (eds) Encyclopedia of Production and Manufacturing Management. Springer, Boston, MA. https://doi.org/10.1007/1-4020-0612-8_874.

[26] R. Koenker, V. Chernozhukov, X. He and L. Peng, (Eds.). (2017). Handbook of quantile regression.

[27] M. Ditzhaus, R. Fried and M. Pauly, QANOVA: quantile-based permutation methods for general factorial designs. TEST, 2021, 30(4), 960-979.

[28] M. Baumeister, M. Ditzhaus and M. Pauly, Quantile-based MANOVA: A new tool for inferring multivariate data in factorial designs. Journal of Multivariate Analysis, 2024, *199*, 105246.